# Scanning Ultrafast Electron Microscopy: A Novel Technique to Probe Photocarrier Dynamics with High Spatial and Temporal Resolutions

Bolin Liao[1*], Ebrahim Najafi[2]

1. Department of Mechanical Engineering, University of California, Santa Barbara, CA, 93106, USA
2. Division of Chemistry and Chemical Engineering, California Institute of Technology, Pasadena, CA, 91125, USA

**Abstract**

The dynamics of photo-excited charge carriers, particularly their transport and interactions with defects and interfaces, play an essential role in determining the performance of a wide range of solar and optoelectronic devices. A thorough understanding of these processes requires tracking the motion of photocarriers in both space and time simultaneously with extremely high resolutions, which poses a significant challenge for previously developed techniques, mostly based on ultrafast optical spectroscopy. Scanning ultrafast electron microscopy (SUEM) is a recently developed photon-pump-electron-probe technique that combines the spatial resolution of scanning electron microscopes (SEM) and the temporal resolution of femtosecond ultrafast lasers. Despite many recent excellent reviews for the ultrafast electron microscopy, we dedicate this article specifically to SUEM, where we review the working principle and contrast mechanisms of SUEM in the secondary-electron-detection mode from a users' perspective and discuss the applications of SUEM to directly image photocarrier dynamics in various materials. Furthermore, we propose future theoretical and experimental directions for better understanding and fully utilizing the SUEM measurements to obtain detailed information about the dynamics of photocarriers. To conclude, we envision the potential of expanding SUEM into a versatile platform for probing photophysical processes beyond photocarrier dynamics.

[*]To whom correspondence should be addressed. Email: bliao@engineering.ucsb.edu.

## I. Introduction

Seeing is believing. The direct visualization of physical, chemical and biological processes has long been an important means to unveil the inner-working principles of complex systems in an accurate and intuitive way. In particular, the development of microscopy has been continuously improving our understanding of the structure and function of matter, culminating in the establishment of aberration-corrected transmission electron microscopes that can routinely resolve atomic details of materials[1–4]. In practice, however, devices do not stay in their static or equilibrium state when they are operating. For example, when converting the solar energy into electricity, photovoltaic cells are excited to a highly non-equilibrium state by absorbing the incoming photons. The photo-excited charge carriers thermalize within themselves and with lattice vibrations, and at the same time transport in the device until they are either collected at the contacts or dissipated via recombination. These fundamental physical processes - generation, recombination and transport, which largely determine the efficiency of a photovoltaic cell, typically occur at timescales that range from femtoseconds to nanoseconds[5]. Therefore, to depict a complete picture of the inner workings of devices, it is insufficient to only look at the static atomic structure of their building materials.

Ultrafast electron microscopy[6–11] is a paradigm-shifting technique that adds nanosecond to sub-picosecond temporal resolution to the conventional electron microscopy. With a probe beam consisting of ultrashort electron pulses generated by illuminating a photocathode with ultrafast laser pulses[12–14], photo-induced dynamic processes[15], such as phase transition[16–18], structural dynamics[19,20] and electromagnetic response[21–24] etc. can be examined in both space and time in either single-shot or stroboscopic modes. Within the family of ultrafast electron microscopy[8], scanning ultrafast electron microscopy (SUEM)[25,26] is a newly developed technique that is particularly suitable for studying photocarrier dynamics in its secondary-electron-detection mode. SUEM interfaces a conventional scanning electron microscope (SEM) with a femtosecond ultrafast laser, and operates on the principle that the local secondary electron yield is related to the local electron or hole population[27,28]. Since it is based on an SEM, SUEM can be used to characterize normal SEM samples, including bulk and microscopic samples with different structures and surface roughness. The capability to measure bulk samples also largely simplifies the sample preparation process and mitigates the sample heating issue compared to the ultrafast transmission electron microscopy. SUEM experiences the same charging issue as the normal SEM, so in principle is not suitable to study electrically insulating materials, although the environmental-mode SUEM[29] can be a potential solution. SUEM has been utilized to image ultrafast photocarrier dynamics on the surface of a wide range of materials, including crystalline semiconductors[28,30], semiconducting nanowires[31] and nanocrystals[32], amorphous semiconductors[33], semiconductor p-n junctions[34] and two-dimensional materials[35], and these applications have resulted in intriguing observations such as ballistic transport of photocarriers across a p-n junction[34], superdiffusion of photocarriers in heavily-doped semiconductors[30] and spontaneous spatial separation of electrons and holes in amorphous semiconductors[33]. Whereas there has been an abundance of recent reviews of ultrafast electron microscopy[8,9,36–38], we dedicate this article specifically to SUEM, with an emphasis on the current understanding of various physical processes that contribute to the contrast images observed in SUEM from a users' perspective. In Section II, we describe the current implementation, the working principle and contrast mechanisms of SUEM in its secondary-electron-detection mode, including our current understanding for data interpretation and its limitations. In Section III, we review and discuss experimental results obtained so far using SUEM on a variety of samples. In

Section IV, we propose future directions in both theory and experiment to further establish the SUEM technique and unleash its full power in understanding photocarrier dynamics, as well as envision the potential of SUEM being expanded into a versatile ultrafast characterization platform by incorporating other commercially available SEM detection modes, such as electron back-scattering diffraction (EBSD). Here we note the parallel development of other techniques that are equally capable of spatial-temporal imaging of photocarrier dynamics with high resolutions, such as time-resolved photoelectron emission microscopy (tr-PEEM)[39–42] and scanning-tip-based near-field methods[43]. These techniques are not reviewed in this article, and interested readers are referred to relevant literature cited here.

**II. Implementation, Working Principle and Contrast Mechanism**

The SUEM prototype described here was designed and implemented by Zewail and coworkers at California Institute of Technology (Caltech) [25], where a conventional SEM (FEI Quanta 650 FEG) was coupled with an ultrafast infrared fiber laser (Clark MXR, 300 fs, 1030 nm). In this setup, as shown in Fig. 1(a), the ultrashort laser pulses generated by the laser source is split into pump and probe beams. The pump beam is directed onto the sample to initiate photophysical processes after frequency conversion for specific experiments. Similarly, the probe beam is typically frequency tripled or quadrupled and focused onto the photocathode (in this case zirconium-oxide coated tungsten filament) inside the SEM column to generate short electron pulses through the photoelectric effect. The time delay between the optical pump pulse and the electron probe pulse is controlled by a mechanical delay line that adds to the optical path of the pump beam. The mechanical delay line typically provides a range of time delays up to a few nanoseconds, which can be potentially expanded by using nanosecond lasers and electronic delay generators[26]. When the probe electron pulse (primary electrons) impacts the sample, the kinetic energy of primary electrons is transferred to electrons within the sample through inelastic scatterings, a portion of which emit to vacuum as secondary electrons, typically from the top few nanometers of the sample surface. Similar to conventional SEM, in SUEM images are formed by scanning the primary electrons across the sample surface and counting the number of emitted secondary electron from each pixel with an Everhart-Thornley detector (ETD)[44], as illustrated in Fig. 1(b). In SUEM, the only difference is that the continuous primary electron beam is replaced by a pulsed beam, and typically the pulsed primary beam dwells at each pixel for up to a few microseconds to allow multiple electron pulses to interact with the sample to obtain sufficient statistics. The excitation of the sample with the optical pump pulse induces electronic or structural changes in the sample, such as photocarrier generation, surface photovoltage, temperature rise, strain and topographical distortions, all of which can potentially enhance or suppress the secondary electron emission. This photo-induced change of secondary electron yield at each pixel and different time delays is the quantity directly measured by SUEM. In practice, a reference image is usually recorded when the electron probe pulse arrives long before the optical pump pulse, when supposedly no photo-induced changes will be observed, assuming the recovering time of the sample is much shorter than the interval of the pulse trains - a prerequisite for SUEM experiments. Then the reference image is subtracted from SUEM images taken at other time delays to remove the static background. The resulting "contrast images", in principle, represent only the photo-induced changes in the sample and their evolution in both space and time.

There are several key parameters that determine the technical performance of SUEM. The temporal resolution of SUEM is determined by the duration of the optical pump and the electron probe pulses, the precision of the mechanical delay stage, and the duration of the secondary electron emission process. The laser pulse width of the setup at Caltech is 300 fs, and the mechanical delay stage has a specified precision of 1 ps. Currently, there is no direct measurement for the duration of the electron pulse and the secondary electron emission process. In practice, the rising time of the SUEM signal from a standard sample, such as intrinsic silicon[27] or cadmium selenide[26], is used to estimate the temporal resolution of this instrument - typically a few picoseconds. It should be noted that the experimental rising time only sets an upper limit for the temporal resolution, since the rising time can also be affected by the vertical transport of photocarriers caused by surface potentials[30]. It is observed that the rising time is affected by the number of electrons per pulse as well[26], which indicates the contribution from the electron pulse width. It is noteworthy here that if the primary electrons are detected instead of the secondary electrons, such as in the electron back-scattering diffraction (EBSD) measurement, the temporal resolution in principle will only be limited by the width of optical and electron pulses, since the interaction time between the primary electrons and the sample becomes much shorter. The spatial resolution of SUEM is related to the electron acceleration voltage and the electron optics. In SUEM, the electron emission filament is illuminated from the side and therefore electrons are emitted not only from the sharp tip region, but also from the main body of the filament. The side illumination likely causes asymmetric electron emission and lowers the spatial coherence of the electron source[45,46] when compared to that of the field emission guns where electrons are drawn out only from the sharp tip. It has been reported that at 30 kV acceleration voltage of the electron beam, a spatial resolution of ~10 nm is achievable with SUEM[25]. The spatial resolution in the contrast images with a given integration time, however, is limited by the signal-noise-ratio (SNR), which is low in the current implementation of SUEM. To compensate for the low SNR, practically a long averaging time (or a large number of images to average over) is required to obtain images with a sufficient quality. One particular noise source is the image intensity fluctuation due to the relative motion between the electron gun and the ultraviolet laser beam that generates the primary electrons.

The contrast images represent the local changes in the secondary electron yield of the sample due to photo-excitation through several possible effects: photocarrier generation, (lattice) temperature rise, strain and topographical distortions. The topographical changes are used to study light-induced mechanical phenomena (e.g. surface acoustic waves) and will not be discussed here. The secondary electron yield is determined by two processes: the generation of secondary electrons inside the sample, and the transport of secondary electrons within the sample before they can escape into the vacuum. Generally, the higher the average energy of the generated secondary electrons, the more of those can escape into vacuum and reach the detector. Photocarrier generation, temperature rise and strain can all affect the average energy of the secondary electrons, while photocarrier generation typically dominates the effect due to its much larger energy scale (a few electronvolts, the photon energy) compared to temperature and strain (tens of milli-electronvolts, set by the thermal energy and the deformation potential). When there is a net accumulation of electrons, the local average electron energy is higher than the equilibrium value, and thus, the local secondary electron yield will increase, resulting in a bright contrast in an SUEM contrast image, a process illustrated in Fig. 2(a). Similarly, a local net accumulation of holes generates a dark contrast. In certain materials, however, the situation is more complicated for two reasons: 1) the scattering of the secondary electrons by photo-excited carriers while they are traveling inside the material and 2) the existence of surface potentials and the surface photovoltage effect. It was previously reported that in intrinsic, n-type and p-type gallium arsenide (GaAs), only a dark contrast emerges in the contrast images in the entire time delay range, shown in Fig. 3(a), implying a reduced secondary electron yield due to photocarriers[28]. This is explained by the increased scattering of secondary electrons by photocarriers that offsets their increased average energy, as schematically shown in Fig. 2(b). A systematic SUEM study of CdSe thin films with different thicknesses suggests[47] that when the sample thickness is thinner than the average mean free path of the secondary electrons, the enhanced secondary electron energy due to the presence of photocarriers dominates and the secondary electron yield increases. On the other hand, if the sample thickness is thicker than the average mean free path of the secondary electrons, the strengthened scattering of secondary electrons by the photocarriers reduces the secondary electron yield. Furthermore, when there are defects and trapping sites on the sample surface, a built-in electrostatic surface potential can be induced. Like the built-in potential at a p-n junction, this surface potential separates electrons and holes, driving one group to the surface and the other to the bulk. This vertical transport process determines whether electrons or holes are observed in SUEM, since the secondary-electron-detection mode is only sensitive to the top few nanometers of the sample, and explains why a bright (dark) contrast is observed in heavily doped p-type (n-type) silicon [30], as shown in Fig. 4 (c). After these electrons and holes are separated by the surface potential, their own electric field can in turn modify the surface potential. This surface photovoltage effect[48] adds to the contrast created by photocarriers since it is dependent on the net photocarrier concentration on the surface. It is clearly seen from this discussion that the contrast mechanism of SUEM is complex and involves many physical processes. This calls for extra caution when interpreting SUEM measurements, but it also shows the potential capability of SUEM to probe all these processes, provided systematic control experiments and detailed modeling work are done in parallel.

**III. Recent Results**

During its early development, SUEM was first used to study the temporal behavior of photocarrier dynamics on material surfaces, namely how the average secondary electron yield within the laser-illuminated area evolves with time before and after photo-excitation. In this mode, SUEM has been applied to study the timescale of photocarrier generation in CdSe[27] and the effect of doping on the photocarrier dynamics in GaAs[28], as shown in Fig. 3. This mode of SUEM provides similar information as other ultrafast optical spectroscopies, but also has the advantage of a much higher surface sensitivity due to typically a shorter secondary electron escaping depth than the optical penetration depth. The surface sensitivity can be further enhanced by reducing the accelerating voltage of the electron beam[47]. This high surface sensitivity enabled, for example, the study of surface states and surface morphology and their effects in photocarrier recombination in indium gallium nitride (InGaN) nanowires[31,49], multinary copper indium gallium selenide (CIGSe) nanocrystals[32] and CdSe[50]. When used in the environmental SEM mode, SUEM can also study photocarrier dynamics on sample surfaces in the presence of water vapor and other gases[29].

Combining the spatial information of photocarrier dynamics with its temporal evolution is a capability of SUEM that goes beyond most optical methods. One exception is the transient absorption microscopy (TAM), an optical pump-probe technique that acquires spatial resolution by scanning one or both of the pump and the probe beams across the sample surface[51–54]. Although it is still limited by optical diffraction, TAM has been used to study photocarrier diffusion in space and time by measuring the temporal change of the spatial size of the photocarrier distribution, the resolution of which is only limited by SNR and can reach tens of nanometers[54]. But in principle TAM cannot resolve more complicated photocarrier dynamics with subwavelength spatial features. In comparison, SUEM is not diffraction-limited and holds the promise of reaching nanometer spatial resolution, although the spatial resolution in contrast images achieved so far has been limited by the SNR. Here we discuss a few examples of the SUEM application with the spatial information for illustrative purpose.

Photocarrier dynamics at a p-n junction is a classical problem in semiconductor physics, and is practically relevant to photovoltaic cells, photodetectors and other optoelectronic devices. Najafi et al. applied SUEM to directly visualize the photo-excitation, electron-hole separation and recombination in space and time at a silicon p-n junction[34]. As shown in Fig. 4(a), the SUEM images taken on the surface of a uniform p-type silicon region show the simple dynamics of the photocarrier generation process. When the photocarriers are generated right at the p-n junction interface, as marked in Fig. 4(b), richer dynamics emerge, including photo-generation (6.7 ps and 36.7 ps), electron-hole separation by the built-in potential within the depletion layer (80 ps), and the gradual recombination of photocarriers (653.3 ps and 3.32 ns). A special characteristic of photocarriers is their initial high temperature and kinetic energy, determined by the photon energy, which lead to the hot carrier transport that can be orders of magnitude faster than normal diffusion. In this case it is observed that the cross-junction transport of hot electrons and holes covers a range of ~80 μm within only ~80 ps, as revealed intuitively in the SUEM contrast images. This hot carrier transport behavior is also observed with SUEM in other materials. For example, on the surface of heavily-doped silicon, two distinct regimes of photocarrier transport are found[30], as shown in Fig. 4(c) and (d): an initial stage of fast expansion up to ~100 ps after photo-excitation, and the transition to a slower normal diffusion regime. It is further found that the initial expansion gets faster with an increasing pump laser fluence, while the transport in the normal diffusion regime is not affected by the pump laser fluence[30]. With a moderate pump laser fluence, the generated photocarriers can be treated as a nondegenerate gas of electrons and holes. While the temperature of the gas is set by the photon energy, the pressure is determined by the number density of photocarriers that depends on the pump laser fluence. The initial fast expansion of the hot carrier gas can be modeled as a result of its initial high temperature and pressure[30], which decrease rapidly with the expansion. These observations illustrate the drastic difference between the hot photocarrier dynamics and the charge transport near equilibrium.

SUEM has also been applied to other types of materials, including amorphous semiconductors[33] and layered van der Waals materials[35]. For example, in hydrogenated amorphous silicon, an archetypical amorphous semiconductor with broad industrial applications, a spontaneous spatial separation of electrons and holes after photo-excitation is observed[33], in addition to a sharp transition from the initial fast diffusion to trapping, when photocarriers cool down below the mobility edge, as shown in Fig. 5 (a) and (b). One convenient measure of the spatial distribution of the photocarriers is the variance of the distribution defined as

$$l^2(t) = \frac{\iint \rho(\mathbf{r},t) r^2 d^2\mathbf{r}}{\iint \rho(\mathbf{r},t) d^2\mathbf{r}} - \left(\frac{\iint \rho(\mathbf{r},t) r d^2\mathbf{r}}{\iint \rho(\mathbf{r},t) d^2\mathbf{r}}\right)^2, \qquad (1)$$

where $\rho(\mathbf{r},t)$ is the image intensity as a function of space and time. The variance is plotted in Fig. 5 (b) to quantitative show the clear diffusion-trapping transition. The spontaneous spatial separation of photo-excited electrons and holes, most clearly seen in the contrast image taken at 827 ps, is caused by the large difference between their mobilities, as well as the low electrical conductivity of the material, which leads to a long dielectric relaxation time that determines how quickly a net charge distribution can be neutralized in a material via electrical conduction. This "relaxation semiconductor" behavior[55] was theoretically predicted in the 1970s, and the SUEM measurement provides direct evidence for its existence in amorphous semiconductors. Another example is black phosphorus, a layered van der Waals material that has attracted increasing interest thanks to its intriguing electrical and optical properties[56,57]. Due to its puckered honeycomb lattice structure[58], black phosphorus has been predicted theoretically to have highly anisotropic charge transport along two in-plane directions[59]. In Fig. 6 (a) and (b), the SUEM images of photo-excited hot holes in black phosphorus clearly verify the strongly anisotropic transport[35]. Plotted in Fig. 6 (b) is the angle-dependent variance of the spatial distribution of hot holes, defined as

$$l^2(\theta, t) = \frac{\iint \rho(\mathbf{r},t)(\mathbf{r}\cdot\hat{\theta})^2 d^2\mathbf{r}}{\iint \rho(\mathbf{r},t) d^2\mathbf{r}} - \left(\frac{\iint \rho(\mathbf{r},t)(\mathbf{r}\cdot\hat{\theta}) d^2\mathbf{r}}{\iint \rho(\mathbf{r},t) d^2\mathbf{r}}\right)^2, \qquad (2)$$

where $\hat{\theta}$ is a unit vector pointing to a certain angle $\theta$. In Fig. 6 (b), the angle-dependent variance is normalized to the initial values and clearly shows that the diffusion of hot holes along the armchair (x) direction is much fast than the zigzag (y) direction. Furthermore, the mobility ratio extracted from the SUEM measurement is around 15, which is much higher than the theoretical prediction and the value measured under near-equilibrium conditions, demonstrating again the distinction between the hot carrier transport and near-equilibrium transport. These examples showcase the capability of SUEM to provide insights to photocarrier dynamics in a wide range of material systems with high spatial and temporal resolutions.

IV. Summary and Future Directions

Being a nascent technique, SUEM has shown great promise of providing detailed information about photocarrier dynamics in distinct material systems, resolved in both space and time with high resolutions, which in turn can be crucial for understanding and designing solar and optoelectronic devices. On the other hand, there are technical details that are currently not sufficiently understood, which can complicate the data interpretation and hinder the utilization of the full capability of SUEM. To make SUEM a quantitative tool, for example, requires knowing the quantitative relationship between the change of secondary electron yield and the local photocarrier concentration and energy. Due to the inherent complexity of second electron emission processes[44], numerical modeling and simulations are indispensable tools for this goal. With the fast-paced advancement of computational material science, we envision that a combined *ab initio* scattering calculation (electron-electron and electron-phonon scatterings) and mesoscopic transport simulation (e.g. Monte Carlo simulation) will be able to fully describe the contrast mechanism of SUEM and help accurately interpret SUEM results. On the experimental side, an improved SNR is desired, which can potentially be achieved by lock-in detection and more stable alignment between the probe laser beam and the electron filament. To single out contributions from specific effects and processes, such as the surface potential and the surface photovoltage effect, systematic control experiments need to be carried out, particularly in conjunction with other surface sensitive characterization techniques, such as Auger electron spectroscopy, low-energy electron diffraction (LEED) and scanning-probe based methods.

Mostly used in secondary-electron-detection mode so far, SUEM can potentially be expanded into a versatile ultrafast characterization platform by incorporating other commercially available SEM detectors. For example, time-resolved EBSD[27] can be used to study photo-excited surface lattice dynamics, particularly suitable for two-dimensional materials and thin films. Recently time-resolved LEED in the transmission mode has been applied to study the photo-induced phase change process of a graphene-polymer composite[18] and ultrafast current in indium phosphide nanowires[60], which can be implemented in SUEM in both reflection and transmission geometries. Time-resolved cathodoluminescence (CL)[61] is a powerful probe for optical properties of nanostructures and can be readily implemented in SUEM.

In summary, the current research status of SUEM is reviewed in this article, including its implementation, working principles and contrast mechanisms as we currently understand, and recent demonstrations of its capability of visualizing photocarrier dynamics in several material systems. Future research directions, including numerical modeling, experimental characterization and further expansion for more capabilities, are also proposed. We envision that SUEM will become a powerful tool for visualizing and understanding photophysical processes on surfaces and interfaces of micro- and nano-scale materials.

**Acknowledgement**

We thank Ding-Shyue Yang for reading and commenting on this manuscript. This work is supported by a start-up fund from the University of California, Santa Barbara.

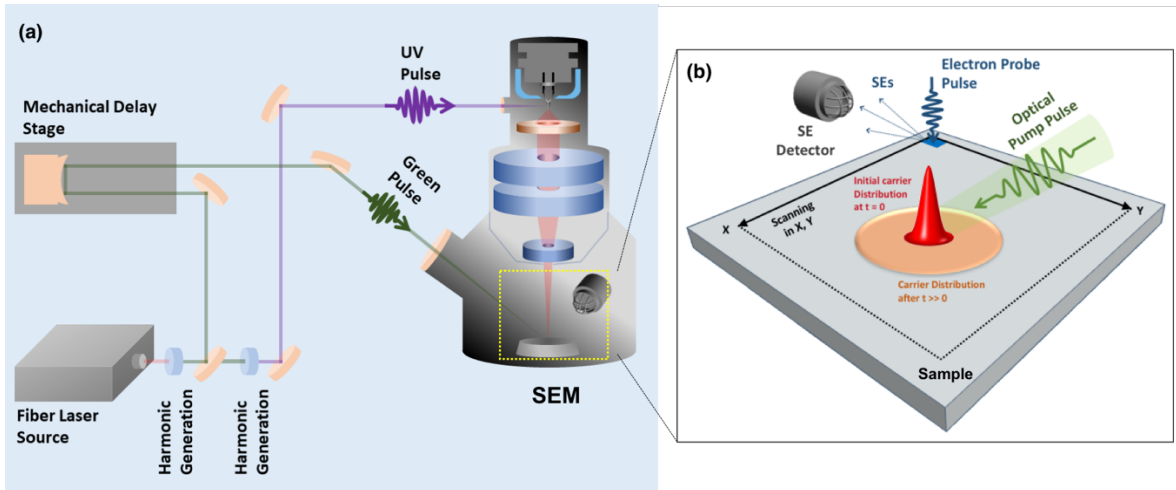

**Figure 1. Schematics of SUEM setup and working principle.** (a) The setup of SUEM. In this case the pump beam is green, and the probe beam is ultraviolet (UV). The green pump beam is focused onto the sample through an optical window on the sample chamber. The UV beam is focused onto the photocathode inside the SEM column to generate ultrashort electron pulses, which are then accelerated and focused onto the sample through the electron optics. (b) The image formation mechanism of SUEM. The electron probe pulse is scanned across the sample surface, and the number of secondary electrons detected from each pixel is used to form an image. The change of the local secondary electron yield caused by the excitation of the optical pump pulse (photocarrier generation in this illustration) is the quantity directly measured by SUEM.

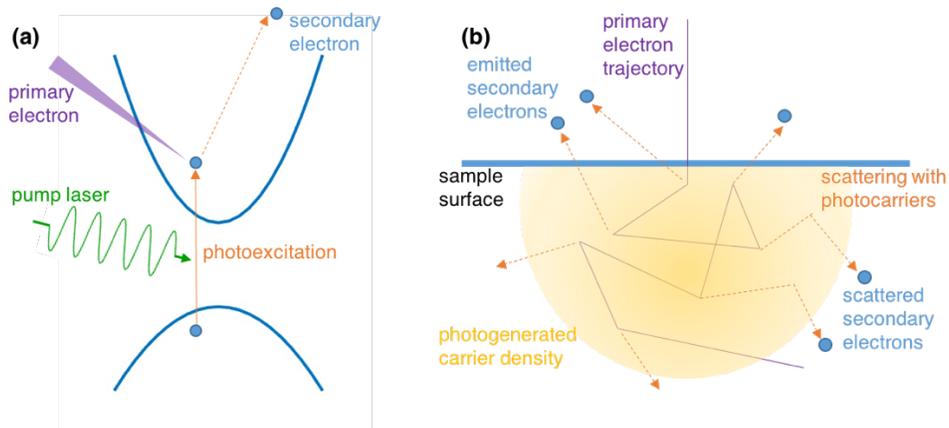

**Figure 2. Schematic showing effects of photocarriers on secondary electrons.** (a) The schematic shows the process of an electron first being excited into the conduction band by absorbing the energy of a pump photon, and then elevated to become a secondary electron after impact of a primary electron. In this process the photo-excitation increases the average energy of generated secondary electrons. (b) The schematic shows a typical trajectory (purple line) of a primary electron as it travels inside the sample, experiencing inelastic scatterings with electrons in the material and generating secondary electrons, which subsequently transport inside the sample (denoted by orange dashed arrows). The yellow shadow denotes the density of charge carriers generated by photo-excitation. Their presence can increase the average energy of secondary electrons as illustrated in (a), but they can also scatter the secondary electrons and prevent them from reaching the sample surface. The two counteracting mechanisms determine the change of secondary electron yield caused by photo-generated charge carriers.

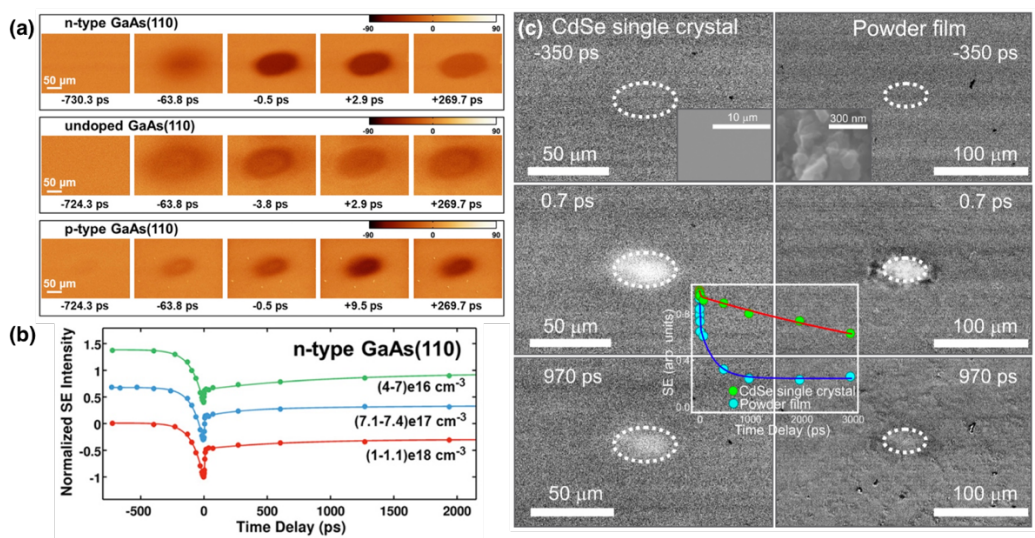

**Figure 3. Application of SUEM for temporal behavior of phocarrier density.** (a) SUEM contrast images on GaAs (110) with different doping conditions. The images are false-colored and the background color denotes the region without the illumination of the pump laser. Only a dark contrast is observed here due to the scattering of secondary electrons by the photo-generated charge carriers, as explained in the main text. (b) The temporal evolution of the average intensity within the laser-illuminated area, indicating the time-scales of photocarrier generation and recombination. (c) SUEM contrast images of CdSe single crystal (left column) and powder film (right column). Inset shows the temporal evolution of the average intensity within the laser-illuminated area, suggesting that photocarriers in the CdSe powder film recombine much faster that those in the CdSe single crystal. (a) and (b) reproduced with permission from reference [28]. © 2014 the National Academy of Sciences of the United States of America. (c) reproduced with permission from reference [50]. © 2016 American Chemical Society.

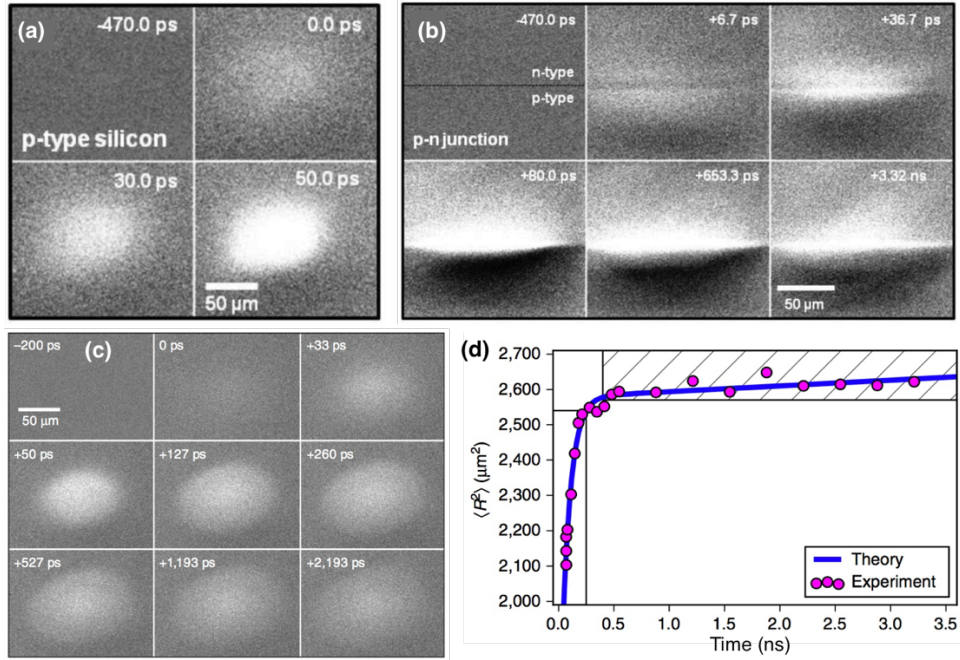

**Figure 4. SUEM imaging of silicon p-n junction and heavily doped silicon.** (a) SUEM contrast images on the p-type region of a silicon p-n junction wafer. (b) SUEM contrast images when photocarriers are generated right at the p-n junction interface. (c) SUEM contrast images showing the initial fast expansion of photocarriers excited on the surface of heavily-doped p-type silicon. (d) The temporal evolution of the variance of the photocarrier spatial distribution extracted from data shown in (c), indicating two distinct regimes of transport with drastically different diffusivities. (a) and (b) reproduced with permission from reference [34]. © 2015 the American Association for the Advancement of Science. (c) and (d) reproduced with permission from reference [30]. © 2017 Nature Publishing Group.

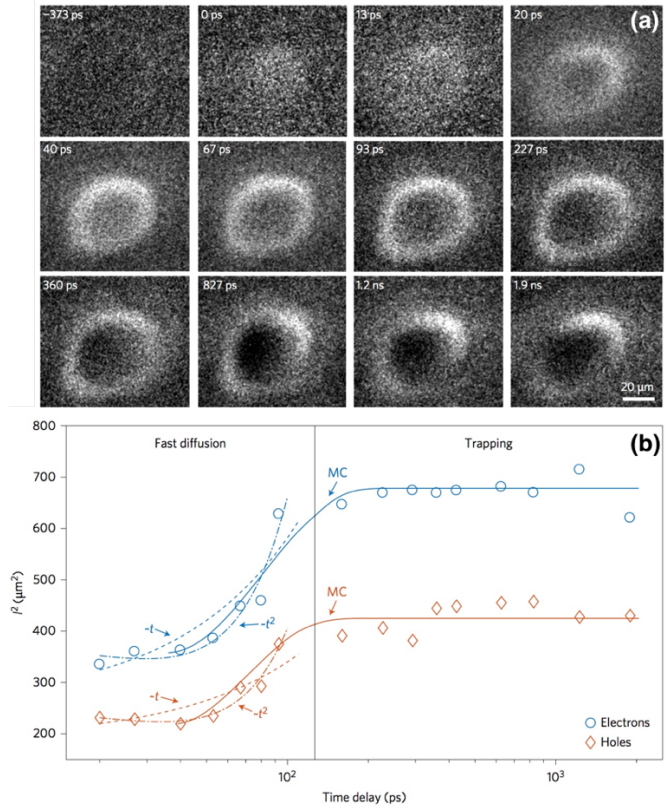

**Figure 5. SUEM imaging of photocarrier dynamics in hydrogenated amorphous silicon.** (a) SUEM contrast images taken on the surface of hydrogenated amorphous silicon, showing the initial fast expansion and the spontaneous spatial separation of hot electrons and holes. (b) The temporal evolution of the variance of spatial distributions of electrons and holes (defined in the main text), as compared to a Monte Carlo (labeled "MC") simulation, and linear (~t) and quadratic (~$t^2$) fits. (a) and (b) reproduced with permission from reference [33]. © 2017 Nature Publishing Group.

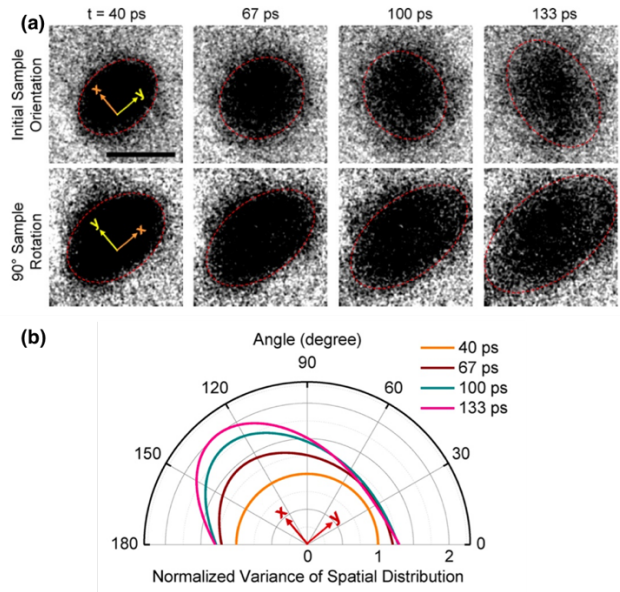

**Figure 6. SUEM imaging of anisotropic hot carrier dynamics in black phosphorus.** (a) SUEM contrast images of hot-holes distribution on the surface of a black phosphorus sample with two orthogonal orientations (the second row corresponds to a 90-degree sample rotation from the first row). "x" denotes the armchair direction and "y" denotes the zigzag direction. It is seen in both cases photo-generated holes preferentially diffuse along the armchair direction. The dashed ellipses are to guide the eye. Scale bar: 60 μm. (b) the temporal evolution of the angle-resolved variance of the hot-holes spatial distribution (defined in the main text), normalized to the initial distribution at 40 ps, quantifying the highly anisotropic diffusion process. (a) and (b) reproduced with permission from reference [35]. © 2017 American Chemical Society.